\documentclass[nofootinbib,aps,prl,twocolumn,showpacs,superscriptaddress]{revtex4-2}

\usepackage{graphicx}
\usepackage{float}
\usepackage{multirow}
\usepackage{amsmath}
\usepackage{comment}
\usepackage{titlesec}
\usepackage{diagbox}
\usepackage{gensymb}
\usepackage[caption=false]{subfig}
\usepackage[utf8]{inputenc}

\usepackage{lineno}

\usepackage{xcolor}

\begin{document}
\title{Limits on the Diffuse Gamma-Ray Background above 10 TeV with HAWC}

\collaboration{HAWC Collaboration}
\author{A.~Albert}
\address{Physics Division, Los Alamos National Laboratory, Los Alamos, NM, USA }
\author{R.~Alfaro}
\address{Instituto de F\'{i}sica, Universidad Nacional Autónoma de México, Ciudad de Mexico, Mexico }
\author{C.~Alvarez}
\address{Universidad Autónoma de Chiapas, Tuxtla Gutiérrez, Chiapas, México}
\author{J.C.~Arteaga-Velázquez}
\address{Universidad Michoacana de San Nicolás de Hidalgo, Morelia, Mexico }
\author{D.~Avila Rojas}
\address{Instituto de F\'{i}sica, Universidad Nacional Autónoma de México, Ciudad de Mexico, Mexico }
\author{H.A.~Ayala Solares}
\address{Department of Physics, Pennsylvania State University, University Park, PA, USA }
\author{R.~Babu}
\address{Department of Physics, Michigan Technological University, Houghton, MI, USA }
\author{E.~Belmont-Moreno}
\address{Instituto de F\'{i}sica, Universidad Nacional Autónoma de México, Ciudad de Mexico, Mexico }
\author{C.~Brisbois}
\address{Department of Physics, University of Maryland, College Park, MD, USA }
\author{K.S.~Caballero-Mora}
\address{Universidad Autónoma de Chiapas, Tuxtla Gutiérrez, Chiapas, México}
\author{T.~Capistrán}
\address{Instituto de Astronom\'{i}a, Universidad Nacional Autónoma de México, Ciudad de Mexico, Mexico }
\author{A.~Carramiñana}
\address{Instituto Nacional de Astrof\'{i}sica, Óptica y Electrónica, Puebla, Mexico }
\author{S.~Casanova}
\address{Institute of Nuclear Physics Polish Academy of Sciences, PL-31342 IFJ-PAN, Krakow, Poland }
\author{O.~Chaparro-Amaro}
\address{Centro de Investigaci\'on en Computaci\'on, Instituto Polit\'ecnico Nacional, M\'exico City, M\'exico.}
\author{U.~Cotti}
\address{Universidad Michoacana de San Nicolás de Hidalgo, Morelia, Mexico }
\author{J.~Cotzomi}
\address{Facultad de Ciencias F\'{i}sico Matemáticas, Benemérita Universidad Autónoma de Puebla, Puebla, Mexico }
\author{S.~Coutiño de León}
\address{Department of Physics, University of Wisconsin-Madison, Madison, WI, USA }
\author{E.~De la Fuente}
\address{Departamento de F\'{i}sica, Centro Universitario de Ciencias Exactase Ingenierias, Universidad de Guadalajara, Guadalajara, Mexico }
\author{R.~Diaz Hernandez}
\address{Instituto Nacional de Astrof\'{i}sica, Óptica y Electrónica, Puebla, Mexico }
\author{B.L.~Dingus}
\address{Physics Division, Los Alamos National Laboratory, Los Alamos, NM, USA }
\author{M.A.~DuVernois}
\address{Department of Physics, University of Wisconsin-Madison, Madison, WI, USA }
\author{M.~Durocher}
\email{mdurocher@lanl.gov}
\address{Physics Division, Los Alamos National Laboratory, Los Alamos, NM, USA }
\author{J.C.~Díaz-Vélez}
\address{Departamento de F\'{i}sica, Centro Universitario de Ciencias Exactase Ingenierias, Universidad de Guadalajara, Guadalajara, Mexico }
\author{K.~Engel}
\address{Department of Physics, University of Maryland, College Park, MD, USA }
\author{C.~Espinoza}
\address{Instituto de F\'{i}sica, Universidad Nacional Autónoma de México, Ciudad de Mexico, Mexico }
\author{K.L.~Fan}
\address{Department of Physics, University of Maryland, College Park, MD, USA }
\author{M.~Fernández Alonso}
\address{Department of Physics, Pennsylvania State University, University Park, PA, USA }
\author{N.~Fraija}
\address{Instituto de Astronom\'{i}a, Universidad Nacional Autónoma de México, Ciudad de Mexico, Mexico }
\author{D.~Garcia}
\address{Instituto de F\'{i}sica, Universidad Nacional Autónoma de México, Ciudad de Mexico, Mexico }
\author{J.A.~García-González}
\address{Tecnologico de Monterrey, Escuela de Ingenier\'{i}a y Ciencias, Ave. Eugenio Garza Sada 2501, Monterrey, N.L., Mexico, 64849}
\author{F.~Garfias}
\address{Instituto de Astronom\'{i}a, Universidad Nacional Autónoma de México, Ciudad de Mexico, Mexico }
\author{M.M.~González}
\address{Instituto de Astronom\'{i}a, Universidad Nacional Autónoma de México, Ciudad de Mexico, Mexico }
\author{J.A.~Goodman}
\address{Department of Physics, University of Maryland, College Park, MD, USA }
\author{J.P.~Harding}
\email{jpharding@lanl.gov}
\address{Physics Division, Los Alamos National Laboratory, Los Alamos, NM, USA }
\author{S.~Hernandez}
\address{Instituto de F\'{i}sica, Universidad Nacional Autónoma de México, Ciudad de Mexico, Mexico }
\author{J.~Hinton}
\address{Max-Planck Institute for Nuclear Physics, 69117 Heidelberg, Germany}
\author{D.~Huang}
\address{Department of Physics, Michigan Technological University, Houghton, MI, USA }
\author{F.~Hueyotl-Zahuantitla}
\address{Universidad Autónoma de Chiapas, Tuxtla Gutiérrez, Chiapas, México}
\author{P.~Hüntemeyer}
\address{Department of Physics, Michigan Technological University, Houghton, MI, USA }
\author{A.~Iriarte}
\address{Instituto de Astronom\'{i}a, Universidad Nacional Autónoma de México, Ciudad de Mexico, Mexico }
\author{V.~Joshi}
\address{Erlangen Centre for Astroparticle Physics, Friedrich-Alexander-Universit\"at Erlangen-N\"urnberg, Erlangen, Germany}
\author{S.~Kaufmann}
\address{Universidad Politecnica de Pachuca, Pachuca, Hgo, Mexico }
\author{D.~Kieda}
\address{Department of Physics and Astronomy, University of Utah, Salt Lake City, UT, USA }
\author{A.~Lara}
\address{Instituto de Geof\'{i}sica, Universidad Nacional Autónoma de México, Ciudad de Mexico, Mexico }
\author{W.H.~Lee}
\address{Instituto de Astronom\'{i}a, Universidad Nacional Autónoma de México, Ciudad de Mexico, Mexico }
\author{H.~León Vargas}
\address{Instituto de F\'{i}sica, Universidad Nacional Autónoma de México, Ciudad de Mexico, Mexico }
\author{J.T.~Linnemann}
\address{Department of Physics and Astronomy, Michigan State University, East Lansing, MI, USA }
\author{A.L.~Longinotti}
\address{Instituto de Astronom\'{i}a, Universidad Nacional Autónoma de México, Ciudad de Mexico, Mexico }
\author{G.~Luis-Raya}
\address{Universidad Politecnica de Pachuca, Pachuca, Hgo, Mexico }
\author{K.~Malone}
\address{Physics Division, Los Alamos National Laboratory, Los Alamos, NM, USA }
\author{O.~Martinez}
\address{Facultad de Ciencias F\'{i}sico Matemáticas, Benemérita Universidad Autónoma de Puebla, Puebla, Mexico }
\author{J.~Martínez-Castro}
\address{Centro de Investigaci\'on en Computaci\'on, Instituto Polit\'ecnico Nacional, M\'exico City, M\'exico.}
\author{J.A.~Matthews}
\address{Dept of Physics and Astronomy, University of New Mexico, Albuquerque, NM, USA }
\author{P.~Miranda-Romagnoli}
\address{Universidad Autónoma del Estado de Hidalgo, Pachuca, Mexico }
\author{J.A.~Morales-Soto}
\address{Universidad Michoacana de San Nicolás de Hidalgo, Morelia, Mexico }
\author{E.~Moreno}
\address{Facultad de Ciencias F\'{i}sico Matemáticas, Benemérita Universidad Autónoma de Puebla, Puebla, Mexico }
\author{M.~Mostafá}
\address{Department of Physics, Pennsylvania State University, University Park, PA, USA }
\author{A.~Nayerhoda}
\address{Institute of Nuclear Physics Polish Academy of Sciences, PL-31342 IFJ-PAN, Krakow, Poland }
\author{L.~Nellen}
\address{Instituto de Ciencias Nucleares, Universidad Nacional Autónoma de Mexico, Ciudad de Mexico, Mexico }
\author{M.~Newbold}
\address{Department of Physics and Astronomy, University of Utah, Salt Lake City, UT, USA }
\author{M.U.~Nisa}
\address{Department of Physics and Astronomy, Michigan State University, East Lansing, MI, USA }
\author{R.~Noriega-Papaqui}
\address{Universidad Autónoma del Estado de Hidalgo, Pachuca, Mexico }
\author{N.~Omodei}
\address{Department of Physics, Stanford University: Stanford, CA 94305–4060, USA}
\author{A.~Peisker}
\address{Department of Physics and Astronomy, Michigan State University, East Lansing, MI, USA }
\author{Y.~Pérez Araujo}
\address{Instituto de Astronom\'{i}a, Universidad Nacional Autónoma de México, Ciudad de Mexico, Mexico }
\author{E.G.~Pérez-Pérez}
\address{Universidad Politecnica de Pachuca, Pachuca, Hgo, Mexico }
\author{C.D.~Rho}
\address{University of Seoul, Seoul, Rep. of Korea}
\author{D.~Rosa-González}
\address{Instituto Nacional de Astrof\'{i}sica, Óptica y Electrónica, Puebla, Mexico }
\author{E.~Ruiz-Velasco}
\address{Max-Planck Institute for Nuclear Physics, 69117 Heidelberg, Germany}
\author{H.~Salazar}
\address{Facultad de Ciencias F\'{i}sico Matemáticas, Benemérita Universidad Autónoma de Puebla, Puebla, Mexico }
\author{F.~Salesa Greus}
\address{Institute of Nuclear Physics Polish Academy of Sciences, PL-31342 IFJ-PAN, Krakow, Poland }
\address{Instituto de F\'{i}sica Corpuscular, CSIC, Universitat de València, E-46980, Paterna, Valencia, Spain }
\author{A.~Sandoval}
\address{Instituto de F\'{i}sica, Universidad Nacional Autónoma de México, Ciudad de Mexico, Mexico }
\author{M.~Schneider}
\address{Department of Physics, University of Maryland, College Park, MD, USA }
\author{J.~Serna-Franco}
\address{Instituto de F\'{i}sica, Universidad Nacional Autónoma de México, Ciudad de Mexico, Mexico }
\author{A.J.~Smith}
\address{Department of Physics, University of Maryland, College Park, MD, USA }
\author{Y.~Son}
\address{University of Seoul, Seoul, Rep. of Korea}
\author{R.W.~Springer}
\address{Department of Physics and Astronomy, University of Utah, Salt Lake City, UT, USA }
\author{O.~Tibolla}
\address{Universidad Politecnica de Pachuca, Pachuca, Hgo, Mexico }
\author{K.~Tollefson}
\address{Department of Physics and Astronomy, Michigan State University, East Lansing, MI, USA }
\author{I.~Torres}
\address{Instituto Nacional de Astrof\'{i}sica, Óptica y Electrónica, Puebla, Mexico }
\author{R.~Torres-Escobedo}
\address{Tsung-Dao Lee Institute and School of Physics and Astronomy, Shanghai Jiao Tong University, Shanghai, China }
\author{R.~Turner}
\address{Department of Physics, Michigan Technological University, Houghton, MI, USA }
\author{F.~Ureña-Mena}
\address{Instituto Nacional de Astrof\'{i}sica, Óptica y Electrónica, Puebla, Mexico }
\author{L.~Villaseñor}
\address{Facultad de Ciencias F\'{i}sico Matemáticas, Benemérita Universidad Autónoma de Puebla, Puebla, Mexico }
\author{X.~Wang}
\address{Department of Physics, Michigan Technological University, Houghton, MI, USA }
\author{E.~Willox}
\address{Department of Physics, University of Maryland, College Park, MD, USA }
\author{H.~Zhou}
\address{Tsung-Dao Lee Institute and School of Physics and Astronomy, Shanghai Jiao Tong University, Shanghai, China }
\author{C.~de León}
\address{Universidad Michoacana de San Nicolás de Hidalgo, Morelia, Mexico }
\author{J.D.~Álvarez}
\address{Universidad Michoacana de San Nicolás de Hidalgo, Morelia, Mexico }

\date{\today}
\begin{abstract}
The high-energy Diffuse Gamma-Ray Background (DGRB) is expected to be produced by unresolved isotropically distributed astrophysical objects, potentially including dark matter annihilation or decay emissions in galactic or extragalactic structures. The DGRB has only been observed below 1 TeV; above this energy, upper limits have been reported. Observations or stringent limits on the DGRB above this energy could have significant multi-messenger implications, such as constraining the origin of TeV-PeV astrophysical neutrinos detected by IceCube. The High Altitude Water Cherenkov (HAWC) Observatory, located in central Mexico at 4100 m above sea level, is sensitive to gamma rays from a few hundred GeV to several hundred TeV and continuously observes a wide field-of-view (2 sr). With its high-energy reach and large area coverage, HAWC is well-suited to notably improve searches for the DGRB at TeV energies. In this work, strict cuts have been applied to the HAWC dataset to better isolate gamma-ray air showers from background hadronic showers. The sensitivity to the DGRB was then verified using 535 days of Crab data and Monte Carlo simulations, leading to new limits above 10 TeV on the DGRB as well as prospective implications for multi-messenger studies.
\end{abstract}

\maketitle

\section{Introduction} 

The Diffuse Gamma-Ray Background (DGRB) is dominated by an isotropic emission of gamma rays uncorrelated with any known sources. The first certain detection of this emission above 50 MeV came from the OSO-3 satellite \citep{OSO} in 1972 and the first spectral measurement above 30 MeV was reported using the SAS-2 satellite in 1975 \citep{SAS}, and later confirmed in 1998 by EGRET \citep{EGRET}. Most hypothesized sources of this diffuse isotropic component to the gamma-ray sky include unresolved active galactic nuclei, starburst galaxies \citep{starburst,DGRB_starburst} and faint gamma-ray bursts \citep{EG-components}. However if the DGRB is observed at multi-TeV energies, where extragalactic emission is attenuated via pair-production on the Extragalactic Background Light (EBL), dark matter annihilation or decay interactions from an extended halo around the Milky Way galaxy may contribute to the diffuse emission \citep{DGRB_DM,EGB_DM}. To fully appreciate the nature of high-energy astrophysical objects it is imperative to understand the origin of the DGRB.

High-energy gamma rays cannot be directly detected from Earth’s surface as they interact with the atmosphere to produce extensive air showers. These air showers consist mainly of relativistic electrons, positrons, and photons. As the average number of particles in the shower increases with depth in the atmosphere, the average energy of each particle decreases until the critical energy is reached and the shower starts to die out \citep{shower_dev}.

This work uses data from the High Altitude Water Cherenkov (HAWC) Gamma-Ray Observatory. The HAWC detector is located in the state of Puebla, Mexico, at an elevation of 4100 meters above sea level. It is sensitive to sources with declinations between -26 and +64 degrees, has a duty cycle of $>$95\% and a wide field-of-view of 2 sr. HAWC uses water as a detection medium, which allows efficient detection of the secondary particles produced during an air shower and calorimetric measurements of the energy deposited in each Water Cherenkov Detector (WCD). With its 300 WCDs, each filled with 200,000 liters of water and containing four photomultiplier tubes (PMTs) anchored to the bottom, HAWC is optimized to detect gamma rays in the energy range between 300 GeV to more than 100 TeV.  In this work, we focus on the energy range between 10 TeV to 200 TeV where HAWC has good gamma/hadron separation capabilities and energy resolution. More information on the design, data acquisition architecture and reconstruction methods of HAWC can be found in \citep{detector1,detector2,detector3}.

The results in this work build upon previous studies performed using the HAWC observatory to set limits on the DGRB \citep{DGRB_pretz,DGRB_zhou,DGRB_harding}. This is achieved by using a bigger area for the DGRB (0.57 sr instead of 0.011 sr), inclusion of estimated energy binning, and improved quantification of the agreement between data and simulation.

\section{Data Selection}

To get strong constraints on the DGRB, we apply tight cuts to remove as much hadronic background as possible from the isotropic gamma-ray signal of interest. The goal for this analysis is to produce a dataset with extremely low background while retaining trust in the agreement between that data and simulation. The remaining gamma-like signal is then compared to simulation to determine its energy and flux. As there are likely some unknown number of hadrons with gamma-like appearance remaining, the results of this study can only be interpreted as upper limits on the DGRB -- a quantification of the expected number of hadronic showers would be needed to declare a positive signal. Additionally, this can only be done if good data/simulation agreement can be demonstrated for these tight cuts. Therefore, the data for this analysis is chosen to be the best quantified data available from HAWC.

\begin{figure*}[hbt!]
\center\includegraphics[width=1\textwidth]{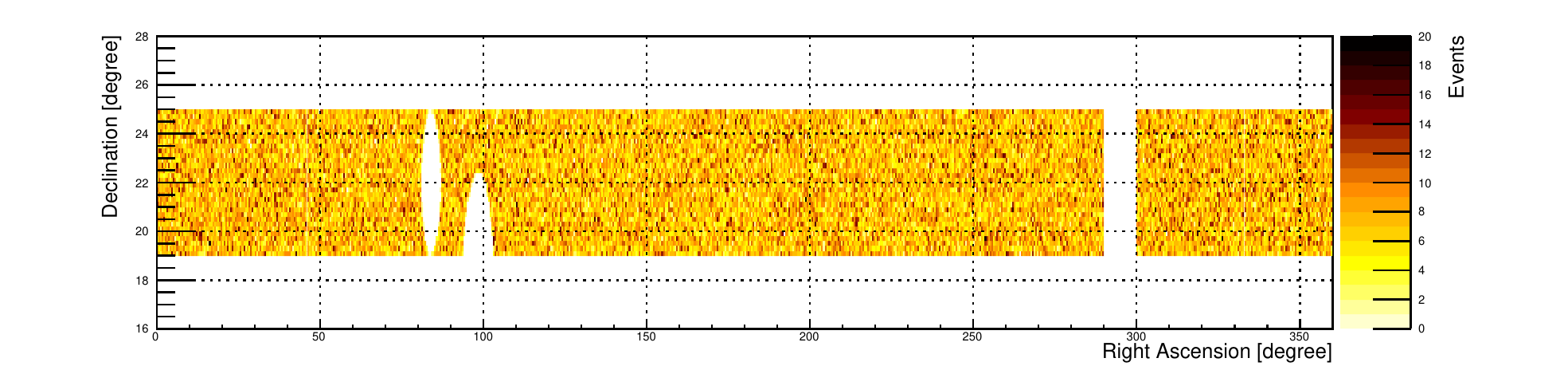}
\caption{Map of the Diffuse Gamma-Ray Background strip, centered on the Crab Nebula's declination. The Crab Nebula, Geminga and the Galactic Plane have been removed, resulting in a 0.57 sr area.} 
\label{fig:DGRB_strip}
\end{figure*}

For our analysis of the DGRB, we focus on a six degrees wide strip centered on the Crab Nebula's declination. As the Crab is a standard reference source in ground-based gamma-ray astronomy, using its location allows us to precisely characterize the performance of the HAWC detector. Focusing on data at the declination of the Crab additionally minimizes any declination-dependent effects which may skew the agreement between data and simulation. To avoid contamination of non-DGRB gamma-ray photons within this strip, bright known gamma-ray sources were removed. Sky locations within $3\degree$ of the Crab itself or within $5\degree$ of the Geminga pulsar-wind nebula or the inner Galactic Plane were excluded from our DGRB search region. To be conservative and avoid any tail, we excluded several Point Spread Functions (PSFs) around these sources. This results in a 0.57 sr area ($\Omega_\text{DGRB}$) we will hereafter refer to as the ``DGRB strip'', which is shown in Figure \ref{fig:DGRB_strip}.

The HAWC data sample used in this work was collected between November 2014 and June 2016, giving a total livetime of 535 days. Although this particular sample does not contain more recent HAWC data, it was chosen for having well characterized gamma/hadron behavior with respect to the HAWC simulation\footnote{This work's gamma-ray simulation relies in part on Corsika v7.4000 \citep{corsika} and GEANT4 v4.10.00 \citep{geant4} to replicate the detector's geometry and performance.}.

In this analysis we implement the same ``ground parameter'' HAWC energy-binning scheme as in \citep{crab} and focus on high energy events, where the HAWC gamma/hadron separation is most efficient.
Starting at 10 TeV, each estimated energy ($\hat{E}$) bin of our analysis spans a quarter decade in log$_{10}$ space, labeled bins $g$ to $l$ (Table \ref{tab:nrg_bin}). Each estimated energy bin is further subdivided into 3 data-quality bins according to the percentage of HAWC PMTs that were triggered during the air shower event. These are numbered as sub-bins $\mathcal{B}$ 7 to 9. This binning structure defines a total of 18 analysis bins, with ranges given in Table~\ref{tab:nrg_bin}.

\begin{table}[h!]
\centering
\begin{tabular}{cc|cc}
\hline 
$\hat{E}$ bin & \text{\hspace{0.5cm}}Energy (TeV) & $\mathcal{B}$ bin & \text{\hspace{0.5cm}}PMTs hit (\%) \\
\hline
$g$ & \text{\hspace{0.5cm}}10.0 - 17.8 & 7 & \text{\hspace{0.5cm}}61.8 - 74.0\\
$h$ & \text{\hspace{0.5cm}}17.8 - 31.6 & 8 & \text{\hspace{0.5cm}}74.0 - 84.0\\
$i$ & \text{\hspace{0.5cm}}31.6 - 56.2 & 9 & \text{\hspace{0.5cm}}84.0 - 100\\
$j$ & \text{\hspace{0.5cm}}56.2 - 100 & &\\
$k$ & \text{\hspace{0.5cm}}100 - 177 & &\\
$l$ & \text{\hspace{0.5cm}}177 - 316 & &\\
\hline 
\end{tabular}
\caption{The HAWC analysis binning. Each $\hat{E}$ bin spans a quarter decade in energy, while the $\mathcal{B}$ sub-bins  define the percentage of HAWC PMTs participating in an event.}
\label{tab:nrg_bin} 
\end{table}
 
\section{Data Quantification}

\subsection{Gamma/Hadron Separation with HAWC Data}

The main background to high-energy photon observation is hadronic cosmic rays; their flux is several orders of magnitude higher than that of gamma rays in the TeV range. Fortunately, above several TeV, the air showers produced by high-energy cosmic rays and gamma rays differ in morphology on the HAWC array. After applying HAWC standard data quality cuts, we employ the gamma/hadron separation parameter called PINCness \citep{pinc} to better isolate gamma-ray showers from cosmic-ray showers. This parameter measures the smoothness of the lateral charge distribution function of air showers as gamma-ray showers have smoother profiles than charged cosmic-ray showers. Requiring lower values of the PINCness removes more hadronic background, though at a cost of removal of greater numbers of true gamma-rays as well. For a given energy bin, we determine the lowest value of PINCness that can be achieved, while retaining a statistically useful number of gamma-like events. This will yield the strongest limits on the DGRB from that bin because the lower PINCness cuts produce higher signal-to-background ratio.

\subsection{Data/Simulation Agreement}
\label{agreement}

To determine how tight a PINCness cut to apply in a given $\hat{E}/\mathcal{B}$ bin, we start with an extremely stringent cut (PINC=0) and loosen it until the number of observed gamma-like events from the Crab Nebula is $\geq2$, which we designate as PINC$_2$. Requiring at least 2 events on the Crab allows us to balance between potential Poissonian fluctuations and setting a very tight cut. To assess whether or not the HAWC data matches the HAWC simulation, we define the Crab data as all events within $0.5\degree$ of the location of the Crab Nebula \footnote{This choice is based on previous HAWC studies of the Crab Nebula \citep{detector2,crab}. In these studies, the radius required to contain 68\% of the photons from the Crab depend on the energy of the shower. In the latest study, the PSF for the Crab varies between $0.25\degree$ and $0.75\degree$. Our choice of events $0.5\degree$ around the Crab Nebula location lies within the range found in said study.}. The simulated events are defined using the best-fit HAWC Crab spectrum \citep{crab}
\begin{eqnarray}
\frac{dN}{dE}\biggr\rvert_\text{Crab}&=&2.35\times10^{-13}\left(\frac{E}{7\text{ TeV}}\right)^{-2.79-0.10\times\text{ln}(E/7\text{ TeV})} \nonumber \\ 
&&\text{[TeV$^{-1}$ cm$^{-2}$ s$^{-1}$]}\enspace
\end{eqnarray}
 
In this regime of strict PINCness cuts, few events remain and Poisson statistics are appropriate to evaluate the level of agreement between data and simulation. We carry out a goodness-of-fit inspection of the PINCness distribution with a binned likelihood analysis. We measure the agreement between data and simulation of the Crab Nebula for all PINCness bins between PINC$_2$ and PINC=1.4 in steps of 0.05 . The maximum PINCness value was chosen based on the tightest gamma/hadron separation cut in the standard HAWC dataset. The log-likelihood is calculated as the sum of the log of the Poisson probability to observe $N^{o,crab}$ events in a bin given that the model predicts $N^{e,crab}$ .
\begin{eqnarray}
\text{ln}\mathcal{L}(N^{e,crab};N^{o,crab})&=&\sum^\text{bins}_{i}N_i^{o,crab}\text{ln}(N_i^{e,crab}) \nonumber \\
&-&\sum^\text{bins}_{i}\left[N_i^{e,crab}+\text{ln}(N_i^{o,crab}!)\right]
\label{eq:log_lik}
\end{eqnarray}
where in each $i$-th PINCness bin, $N^{o,crab}$ is the number of observed events in the Crab Region of interest (RoI) and $N^{e,crab}$ is the number of expected events in that same region i.e. the simulated gamma events in the Crab RoI plus the observed background from data.

A $\chi^2/2$ distributed function based on this binned maximum likelihood analysis is then defined as
\begin{equation}
\Lambda=\text{ln}\cfrac{\mathcal{L}(N^{e,crab};N^{o,crab})}{\mathcal{L}(N^{o,crab};N^{o,crab})}
\end{equation}
where $\text{ln}\mathcal{L}(N^{o,crab};N^{o,crab})$ is the maximum value for the log-likelihood given the observed counts. The resulting p-value ($P_\Lambda$) is used to quantify the level of agreement between data and simulation. 

\begin{figure}[h!]
\center\includegraphics[width=0.45\textwidth]{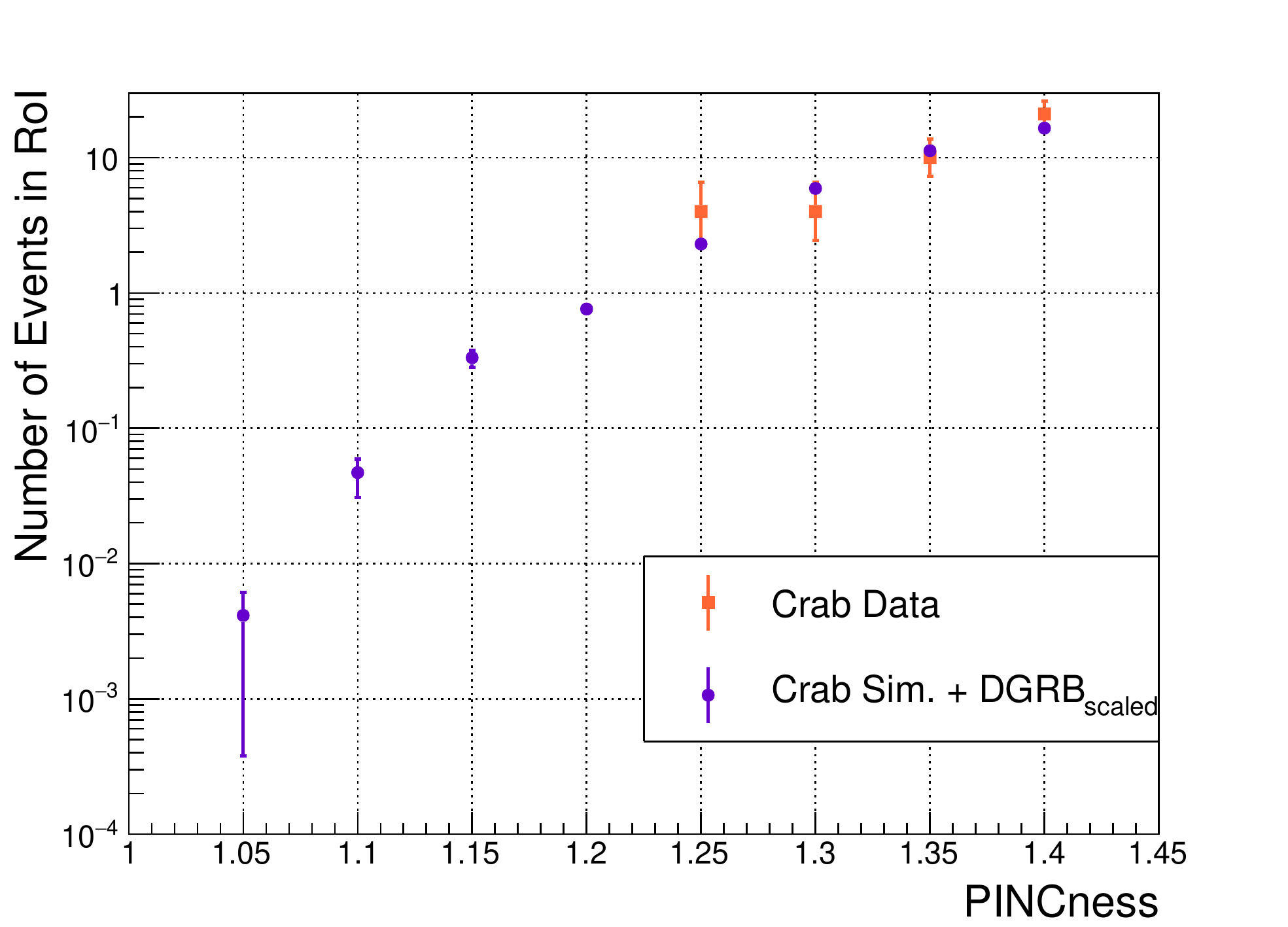}
\caption{Number of observed and expected events in the Crab RoI with respect to PINCness for the $\hat{E}/\mathcal{B}$ bin $g$/7. The expected number of events is the sum of the simulated gamma events from the Crab plus the $\text{DGRB}_\text{scaled}$, which is the average number of events from the DGRB strip found in  $2.22\times10^{-4}$ sr (the size of the Crab RoI).} 
\label{fig:pinc_diff}
\end{figure}

If a given analysis bin's $P_\Lambda>0.32$, we assume a good fit between its Crab data and simulation. We then rely on simulated events down to tighter PINCness cuts at which there is insufficient data on the Crab ($<2$ events from the Crab RoI). This extrapolation is based on the assumption that had the data not run out in this analysis bin, the data/simulation agreement would have persisted. Figure \ref{fig:pinc_diff} illustrates the case of bin $g$/7, where $P_\Lambda=0.56$.

The goal is to find the lowest PINCness cut with good data/simulation agreement. To do so, we make the following requirements:

\begin{enumerate}
\item For analysis bins with $P_\Lambda>0.32$:
\begin{enumerate}
\item $\geq25$ simulated Corsika events
\item $\geq9$ observed events from the DGRB strip
\end{enumerate}
\item For analysis bins with $P_\Lambda\leq0.32$:
\begin{enumerate}
\item $\geq25$ simulated Corsika events
\item $\geq9$ observed events from the DGRB strip
\item $\geq2$ observed events from the Crab RoI
\item Pearson's $\chi^2(N^{e,crab};N^{o,crab})$\\$=\cfrac{(N^{o,crab}-N^{e,crab})^2}{N^{e,crab}}\leq1$
\end{enumerate}
\end{enumerate}
For criteria (a) and (b), we consider the total number of events surviving a given PINCness cut. Criterion (d) is there to verify that there is still good data/simulation agreement on the Crab between $N^{e,crab}$ and $N^{o,crab}$ despite the poor p-value from the distributions. Observations of the Crab Nebula above 50 TeV are sparse as its flux is small in this high energy range and not all analysis bins are populated. For high-energy bins which have insufficient Crab data to calculate $\Lambda$, we also use criteria (a) and (b) to determine the lowest valid PINCness cut.

In this way, we have established the tightest reliable gamma/hadron separation cuts in each HAWC analysis bin. In Table \ref{tab:bin_pinc_mc}, we display the optimal PINCness cut for each analysis bin. This selection will be applied throughout the rest of this work.

\begin{table}[h!]
\centering
\begin{tabular}{c | c c c c c c}
\hline 
\backslashbox{$\mathcal{B}$ bin}{$\hat{E}$ bin} & \text{\hspace{0.4cm}}$g$\text{\hspace{0.4cm}} & \text{\hspace{0.4cm}}$h$\text{\hspace{0.4cm}} & \text{\hspace{0.4cm}}$i$\text{\hspace{0.4cm}} & \text{\hspace{0.4cm}}$j$\text{\hspace{0.4cm}} & \text{\hspace{0.4cm}}$k$\text{\hspace{0.4cm}} & $l$ \\
\hline
7 & 1.15 & 1.1 & 1.25 & --- & --- & --- \\
8 & 1.2 & 1.25 & 1.2 & 1.4 & --- & --- \\
9 & 1.3 & 1.3 & 1.2 & 1 & 1 & 1.05 \\
\hline
\end{tabular}
\caption{The optimal PINCness cut for each analysis bin. Note that bins with optimal PINCness cuts  greater than 1.4 were excluded from this analysis.}
\label{tab:bin_pinc_mc} 
\end{table}

\section{Limits on the DGRB}

\subsection{Differential Flux Limits}

We have established which $\hat{E}/\mathcal{B}$ bins to use and the corresponding optimal PINCness cuts. Now we will calculate differential flux limits as a function of energy. To do so, each analysis bin is treated independently. The log-likelihood $\text{ln}\mathcal{L}(N^{sim};N^{obs})$ for a bin is expressed as
\begin{equation}
N^{obs}\text{ln}(N^{sim})-N^{sim}-\text{ln}(N^{obs}!)
\label{eq:ind_lik}
\end{equation}
where $N^{obs}$ is set as the number of events in our DGRB strip and $N^{sim}$ depends on the number of simulated events. To calculate $N^{sim}$, we inject an E$^{-2.5}$ spectrum in the HAWC simulation and apply the gamma/hadron separation cuts from Table \ref{tab:bin_pinc_mc}. We chose a baseline spectrum of E$^{-2.5}$ as it lies within the range of the astrophysical models that will be presented in \emph{Constraining Astrophysical Models}. To maximize the likelihood that the $N^{sim}$ described by our simulation produced the DGRB data $N^{obs}$, we look for the minimum of $-2\text{ln}\mathcal{L}$ in each analysis bin and calculate the 95\% one-sided upper limit ($2\Delta\text{ln}\mathcal{L}=2.71$). Finally, for a given energy bin, we use the $\mathcal{B}$ bin which produces the smallest upper limit.

A scale factor $\beta$ is multiplied by the differential flux
\begin{eqnarray}
\frac{d^2N}{dEd\Omega}&=&\frac{1}{\Omega_\text{DGRB}}\times10^{-11}\left(\frac{E}{\text{ TeV}}\right)^{-2.5} \nonumber \\
&&\text{[TeV$^{-1}$ cm$^{-2}$ s$^{-1}$ sr$^{-1}$]}
\label{eq:beta_flux}
\end{eqnarray}
with the corresponding 95\% Confidence Level (CL) upper limit referred to as $\beta_{95\%}$ . For this flux, as well as those in Equations \ref{eq:IC_flux} to \ref{eq:LAT_flux}, we will be constraining the normalization of the differential flux through constraints of this multiplication factor.
 
\begin{figure}[h!]
\center\includegraphics[width=0.45\textwidth]{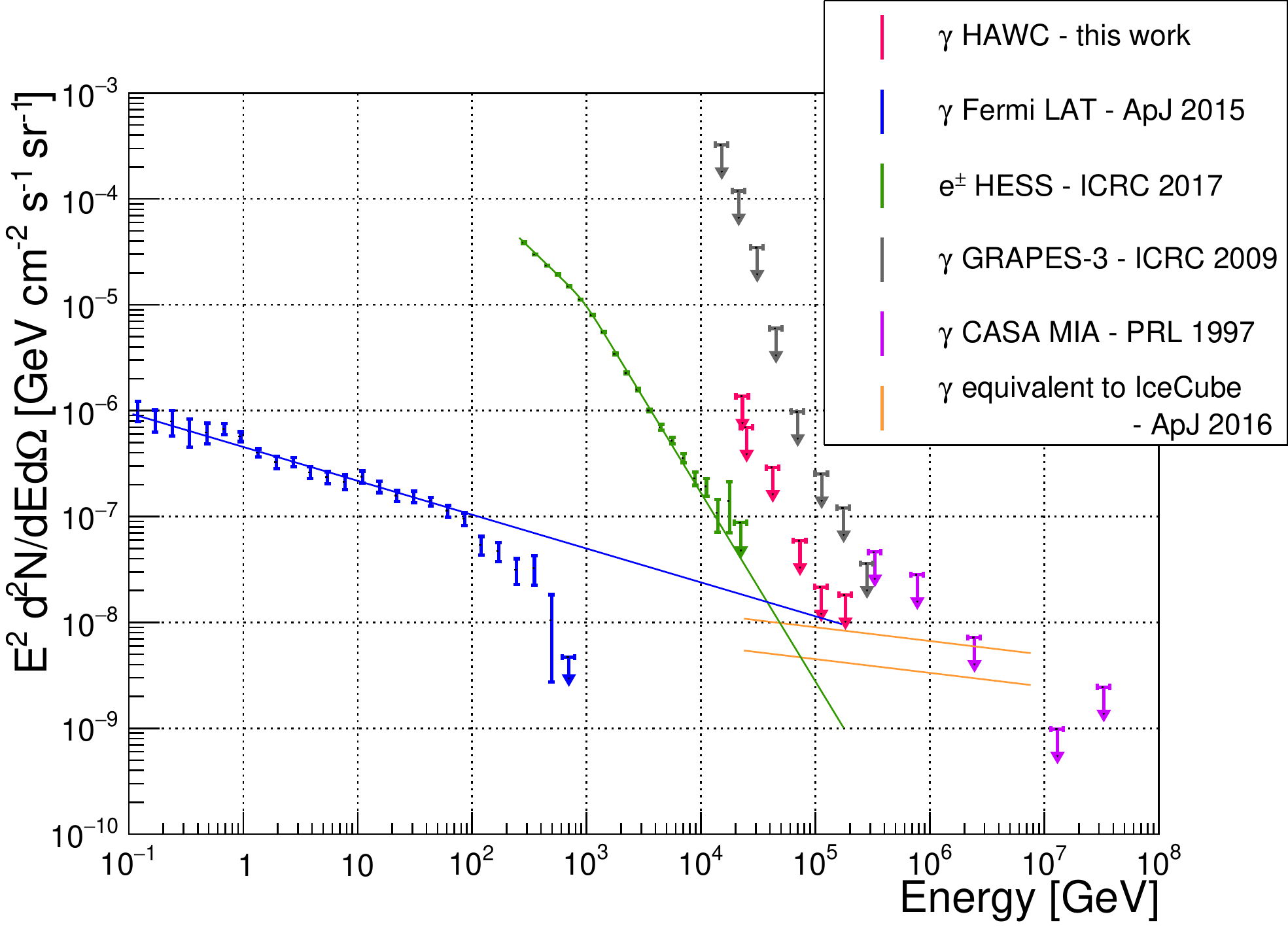}
\caption{Limits on the DGRB using 535 days of HAWC data compared to the diffuse electron/positron flux observed by HESS \citep{hess1_graph,hess2_graph}. Also shown is the observed DGRB by the Fermi-LAT \citep{fermi_graph}, as well as previous high-energy limits by GRAPES \citep{grapes_graph} and CASA-MIA \citep{casa-mia_graph}. The lines represent several astrophysical models which will be discussed in \emph{Constraining Astrophysical Models}. The gamma-ray flux corresponding to the IceCube $\nu_\mu+\overline{\nu}_\mu$ astrophysical flux \citep{icecube} is shown (see \emph{Constraining Astrophysical Models - IceCube Neutrino Spectrum}) for both $p\gamma$ and $pp$ interactions.}
\label{fig:ul_mc}
\end{figure}

For each $\hat{E}$ bin studied, Table \ref{tab:ul_nrg_mc} shows the $\mathcal{B}$ bin with the smallest value of $\beta_{95\%}$ as well as the median energy from simulation, assuming the E$^{-2.5}$ spectrum. Figure \ref{fig:ul_mc} shows the upper limits for each analysis bin compared to other observations and limits.

\begin{table}[h!]
\centering
\begin{tabular}{c c c}
\hline 
$\hat{E}/\mathcal{B}$ bin & \text{\hspace{0.5cm}}$\beta_{95\%}$\text{\hspace{0.5cm}} & Sim. median energy (TeV) \\
\hline
$g$/7 & 657 & 23 \\
$h$/8 & 349 & 25 \\
$i$/9 & 189 & 42 \\
$j$/9 & 50.6 & 73 \\
$k$/9 & 22.9 & 112 \\
$l$/9 & 24.6 & 182 \\
\hline
\end{tabular}
\caption{HAWC analysis bins in which the smallest upper limits were found. The corresponding $\beta_{95\%}$ and median energy from simulation with an E$^{-2.5}$ spectrum are also shown.}
\label{tab:ul_nrg_mc} 
\end{table}

\subsection{Quantifying Uncertainties}

Systematic uncertainties have been derived by computing the Crab spectrum under varying assumptions in the modeling of the detector. The stability of the results in various detector models are studied by altering our detector performance and observing the impact on our limits. More information on the different effects investigated can be found in \citep{crab}. The largest source of uncertainty is related to the magnitude and timing of the light in the HAWC PMTs. This can affect our results by up to a factor of four.

We also consider the effects of varying our power law from E$^{-2.5}$ to E$^{-2}$ or E$^{-3}$. For bins $h$/8 to $l$/9, changing to an E$^{-2}$ spectrum increases gamma energy by 8.4\% but decreases flux by 4.2\% . An E$^{-3}$ spectrum lowers gamma energy by 9.3\% but increases flux by 8.3\%. For bin $g$/7, the change to an E$^{-2}$ spectrum keeps gamma energy within the previous range but increases flux by 9.7\%; while an E$^{-3}$ spectrum lowers gamma energy by 21.7\% but keeps flux within the previous range.

Finally, to assess statistical uncertainty with the E$^{-2.5}$ spectrum, we change the $N^{obs}$ events in the DGRB strip by $\pm1\sigma$ (Poisson fluctuation) and find our results change by less than 30\% .

\section{Constraining Astrophysical Models}
\label{models}

We now evaluate the data in the DGRB strip with respect to the spectra from several astrophysical models. To do so, we employ the same analysis bins as those obtained in \emph{Data Quantification: Data/Simulation Agreement}, as well as a similar analysis method but for astrophysical models extrapolated to HAWC energies.

\subsection{IceCube Neutrino Spectrum}

In the case where the IceCube flux is diffuse from $p\gamma$ or $pp$ interactions, gamma rays are expected to exhibit a similar spectrum to the neutrinos seen by IceCube. Assuming these messengers have a common hadronic origin, we can make use of a relation between the fluxes of gamma rays ($F_\gamma$) and neutrinos ($F_{\nu_\alpha}$) from galactic sources \citep{gamma_nu,amon}
\begin{equation}
E_\gamma F_\gamma(E_\gamma)\approx\frac{2}{3K}\sum_{\nu_\alpha}E_\nu F_{\nu_\alpha}(E_\nu)
\label{eq:gamma_nu}
\end{equation}
where $E_\gamma\approx2E\nu$ and $K$ is the ratio of charged to neutral pions with $K=1$ for photo-hadronic interactions and $K=2$ for hadro-nuclear interactions.

The gamma-ray flux corresponding to the IceCube unbroken power-law model for the $\nu_\mu+\overline{\nu}_\mu$ astrophysical flux \citep{icecube} is then
\begin{eqnarray}
\frac{d^2N}{dEd\Omega}\biggr\rvert_\text{IC,$\gamma$}&=&\frac{1}{K}\times0.90\times10^{-15}\left(\frac{E_\nu}{100\text{ TeV}}\right)^{-2.13} \nonumber \\
&&\text{[TeV$^{-1}$ cm$^{-2}$ s$^{-1}$ sr$^{-1}$]}
\label{eq:IC_flux}
\end{eqnarray}
Using the spectrum of Equation \ref{eq:IC_flux} in our HAWC simulation, the analysis bin whose limit is closest to this model’s best fit extrapolation is $\hat{E}/\mathcal{B}$ bin $l$/9. Its DGRB limit is 2.13 (4.26) times the associated IceCube flux for photo-hadronic (hadro-nuclear) interactions, and the corresponding median energy is 190 TeV.

\subsection{H.E.S.S. Electron/Positron Spectrum}

The H.E.S.S. observatory has detected an isotropic CR flux of electrons and positrons up to 20 TeV \citep{hess1_graph,hess2_graph}. Due to the similar morphology of air showers induced by gamma rays and electrons/positrons, the HAWC detector cannot distinguish between the two. Therefore any HAWC limit on the DGRB is also an upper limit on the isotropic CR flux of electrons and positrons. Although the limits obtained in this work are at an energy range higher than the H.E.S.S. observations, a comparison with the best fit for the H.E.S.S. electron/positron flux with a smooth broken power law can still be extrapolated above 20 TeV
\begin{eqnarray}
\frac{d^2N}{dEd\Omega}\biggr\rvert_\text{HESS}&=&1.05\times10^{-8}\left(\frac{E}{1\text{ TeV}}\right)^{-3.04} \nonumber \\
&&\times\left(1+\left(\frac{E}{0.94\text{ TeV}}\right)^{\frac{1}{0.12}}\right)^{0.12\times(3.04-3.78)} \nonumber \\
&&\text{[TeV$^{-1}$ cm$^{-2}$ s$^{-1}$ sr$^{-1}$]}
\label{eq:HESS_flux}
\end{eqnarray}
Using the spectrum of Equation \ref{eq:HESS_flux} in our HAWC simulation, our most constraining limit is a factor of 9.60 higher than this model’s best fit extrapolation to HAWC’s energy range and is found in $\hat{E}/\mathcal{B}$ bin $k$/9 at a median energy of 103 TeV.

\subsection{Fermi-LAT Gamma-Ray Spectrum}

The isotropic diffuse gamma-ray background flux has been measured by the Fermi-LAT from 100 MeV to 820 GeV \citep{fermi_graph}. In that paper, several model shapes were fitted to the Fermi-LAT data, including a Broken Power Law (BPL) and a Power Law with an Exponential cutoff (PLE). In the BPL and PLE models, the flux expected at the energy range of our analysis is many orders of magnitude lower than the limits we can achieve. However we can get reasonable constraints on the Fermi-LAT simple Power Law (PL) model, which is of the form
\begin{eqnarray}
\frac{d^2N}{dEd\Omega}\biggr\rvert_\text{LAT,PL}&=&0.95\times10^{-1}\left(\frac{E}{100\text{ MeV}}\right)^{-2.32}\nonumber \\
&&\text{[TeV$^{-1}$ cm$^{-2}$ s$^{-1}$ sr$^{-1}$]}
\label{eq:LAT_flux}
\end{eqnarray}
While noting that the Fermi-LAT PL model is disfavored by their data, their PL extrapolation to HAWC energies is only factor of 1.91 below our most constraining limit. This result was obtained in $\hat{E}/\mathcal{B}$ bin $l$/9, at a median energy of 186 TeV.

\subsection{Summary}

We arrange these results in Table \ref{tab:model_mc}, with the corresponding simulation median energy. This summary table shows the overall most constraining limit in the form of the smallest $\beta_{95\%}$. The latter being the 95\% CL upper limit of the scale factor for each astrophysical model.

\begin{table}[h!]
\centering
\begin{tabular}{c c c}
\hline 
Model & \text{\hspace{0.5cm}}$\beta_{95\%}$\text{\hspace{0.5cm}} & Sim. median energy (TeV) \\
\hline
IceCube (Eq. \ref{eq:IC_flux}) & 2.13 & 190 \\
H.E.S.S (Eq. \ref{eq:HESS_flux}) & 9.60 & 103 \\
Fermi-LAT (Eq. \ref{eq:LAT_flux}) & 1.91 & 186 \\
\hline 
\end{tabular}
\caption{Summary of the multi-messenger studies containing the smallest $\beta_{95\%}$ and the median energy from simulation of the astrophysical spectra.}
\label{tab:model_mc} 
\end{table}

\section{Discussion and Prospects}

We applied gamma/hadron separation methods and enforced data/simulation agreement while maximizing the signal-to-background ratio in order to calculate 95\% upper limits on the DGRB. In addition to DGRB events, the flux observed by HAWC may also include other isotropic events that would form an irreducible background. One possible component would be misidentified hadrons, due to their gamma-like appearance in the HAWC detector. This can happen if a proton interacts in the atmosphere to produce mainly neutrinos, electrons, positrons, and gamma rays. This can also happen if high-transverse-momentum parts of the shower, e.g. muons, land outside the detector array. Another possible source of isotropic emission would be cosmic-ray electrons and positrons, whose air showers are indistinguishable from those induced by gamma rays. A better understanding of the expected cosmic-ray contamination would help break the degeneracy and may allow for the calculation of best fit photon flux, and not just upper limits.

It is possible for diffuse gamma-ray emissions to originate from annihilating or decaying dark matter in galactic or extragalactic structures \citep{DGRB_DM,EGB_DM}. Gamma rays would then be observed in all directions, as a background to all other gamma-ray observations. The search for dark matter using this method will be discussed in a future publication.

Astrophysical pion decay produces neutrinos as well as gamma rays, and so might dark matter annihilations. Interestingly, our limit on the IceCube flux is nearly twice as high as said flux for photo-hadronic interactions. Therefore we cannot yet confirm the presence of a diffuse neutrino population using gamma rays. Note that this is valid under the assumption that gamma-ray emissions are not attenuated by the EBL. Better characterization of misindentified cosmic rays is underway and could verify a consistency between the gamma-ray and neutrino flux.

August 2018 marked the completion of an array of 350 outrigger WCDs surrounding the central detector \citep{outriggers}. Each outrigger WCD contains 2000 liters of water and is instrumented with a single PMT. The outriggers are designed to extend the effective collection area of HAWC by about a factor of four for high-energy observations, which will lead to an improvement in shower core and energy resolution by at least a factor of two. Moreover, other experiments with much bigger detection areas such as the next-generation Southern Wide-field Gamma-ray Observatory (SWGO)\footnote{\text{www.swgo.org}} and the Large High Altitude Air Shower Observatory (LHAASO) \citep{LHAASO} will extend sensitivities to energies in the PeV.

Furthermore, notable improvements to the HAWC reconstruction algorithms are in progress. They are expected to reduce the hadronic background by an order of magnitude for the same number of signal events at high energies. With these upcoming improvements and the addition of recent years of data, including the outrigger array, stronger HAWC constraints can be expected in the future.
\begin{acknowledgements}
We acknowledge the support from: the US National Science Foundation (NSF); the US Department of Energy Office of High-Energy Physics; the Laboratory Directed Research and Development (LDRD) program of Los Alamos National Laboratory; Consejo Nacional de Ciencia y Tecnolog\'ia (CONACyT), M\'exico, grants 271051, 232656, 260378, 179588, 254964, 258865, 243290, 132197, A1-S-46288, A1-S-22784, c\'atedras 873, 1563, 341, 323, Red HAWC, M\'exico; DGAPA-UNAM grants IG101320, IN111716-3, IN111419, IA102019, IN110621, IN110521; VIEP-BUAP; PIFI 2012, 2013, PROFOCIE 2014, 2015; the University of Wisconsin Alumni Research Foundation; the Institute of Geophysics, Planetary Physics, and Signatures at Los Alamos National Laboratory; Polish Science Centre grant, DEC-2017/27/B/ST9/02272; Coordinaci\'on de la Investigaci\'on Cient\'ifica de la Universidad Michoacana; Royal Society - Newton Advanced Fellowship 180385; Generalitat Valenciana, grant CIDEGENT/2018/034; The Program Management Unit for Human Resources \& Institutional Development, Research and Innovation, NXPO (grant number B16F630069);; Coordinaci\'on General Acad\'emica e Innovaci\'on (CGAI-UdeG), PRODEP-SEP UDG-CA-499; Institute of Cosmic Ray Research (ICRR), University of Tokyo, H.F. acknowledges support by NASA under award number 80GSFC21M0002. We also acknowledge the significant contributions over many years of Stefan Westerhoff, Gaurang Yodh and Arnulfo Zepeda Dominguez, all deceased members of the HAWC collaboration. Thanks to Scott Delay, Luciano D\'iaz and Eduardo Murrieta for technical support.
\end{acknowledgements}

\bibliography{biblio}
\bibliographystyle{apsrev4-1}

\end{document}